# Lung and Colon Cancer Histopathological Image Dataset (LC25000)

Andrew A. Borkowski, MD*[1,2], Marilyn M. Bui, MD, PhD[2,3], L. Brannon Thomas, MD, PhD[1,2], Catherine P. Wilson, MT[1], Lauren A. DeLand, RN[1], Stephen M. Mastorides, MD[1,2]

[1] Pathology and Laboratory Service, James A. Haley Veterans' Hospital, Tampa, Florida, USA
[2] Department of Pathology and Cell Biology, University of South Florida, Tampa, Florida, USA
[3] Department of Pathology and Analytic Microscope Core, Moffitt Cancer Center, Tampa, Florida, USA

*E-mail: andrew@usf.edu

**Abstract**

The field of Machine Learning, a subset of Artificial Intelligence, has led to remarkable advancements in many areas, including medicine. Machine Learning algorithms require large datasets to train computer models successfully. Although there are medical image datasets available, more image datasets are needed from a variety of medical entities, especially cancer pathology. Even more scarce are ML-ready image datasets. To address this need, we created an image dataset (LC25000) with 25,000 color images in 5 classes. Each class contains 5,000 images of the following histologic entities: colon adenocarcinoma, benign colonic tissue, lung adenocarcinoma, lung squamous cell carcinoma, and benign lung tissue. All images are de-identified, HIPAA compliant, validated, and freely available for download to AI researchers.

## 1. Introduction

The field of Artificial Intelligence (AI) is rapidly growing. Machine Learning (ML), a subset of AI, has the potential for numerous applications in the healthcare fields.[1][2] One promising application is in the field of diagnostic pathology.[3][4] ML allows representative images to be used to train a computer to recognize patterns from labeled photographs. Based on a set of images selected to represent a specific tissue or disease process, the computer can be trained to evaluate and recognize new and unique images from patients and render a diagnosis.[5]

Machine Learning requires large image datasets for training. Although few such datasets are available to researchers, more freely available datasets are needed.[6] To fill this need, we created a color image dataset (LC25000) of benign and cancerous lung and colon tissue images. Carcinomas of the lung and colon are among the most common sources of invasive cancer and are the two most common causes of cancer deaths in America.[7] Improving cancer diagnosis through ML algorithms would hopefully improve these grave statistics.

## 2. Dataset

### 2.1 Image acquisition

HIPAA compliant and validated seven hundred fifty total images of lung tissue (250 benign lung tissue, 250 lung adenocarcinomas, and 250 lung squamous cell carcinomas) and 500 total images of colon tissue (250 benign colon tissue and 250 colon adenocarcinomas) were captured from pathology glass slides as we previously described.[8]

*2.2 Image augmentation*

All images were cropped to square sizes of 768 x 768 pixels from original 1024 x 768 pixels using python programming language. Subsequently, images were augmented using the Augmentor software package. Augmentor is an image augmentation library in Python for machine learning. It aims to be a standalone library that is platform and framework independent, which is more convenient, allows for finer-grained control over augmentation, and implements the most real-world relevant augmentation techniques. It employs a stochastic approach using building blocks that allow for operations to be pieced together in a pipeline.[9]

Using Augmentor, we expanded our dataset to 25,000 images by the following augmentations: left and right rotations (up to 25 degrees, 1.0 probability) and by horizontal and vertical flips (0.5 probability).

*2.3 Dataset description*

The dataset contains 25,000 color images with five classes of 5,000 images each. All images are 768 x 768 pixels in size and are in jpeg file format. Our dataset can be downloaded as a 1.85 GB zip file LC25000.zip.[10] After unzipping, the main folder lung_colon_image_set contains two subfolders: colon_image_sets and lung_image_sets. The subfolder colon_image_sets contains two secondary subfolders: colon_aca subfolder with 5,000 images of colon adenocarcinomas and colon_n subfolder with 5,000 images of benign colonic tissues. The subfolder lung_image_sets contains three secondary subfolders: lung_aca subfolder with 5,000 images of lung adenocarcinomas, lung_scc subfolder with 5,000 images of lung squamous cell carcinomas, and lung_n subfolder with 5,000 images of benign lung tissues.

## 3. Discussion

The field of Machine Learning, a subset of AI, has led to advancements in many fields, including medicine. Numerous studies utilizing ML have been performed in the areas of dermatology, ophthalmology, radiology and pathology.[1][2] ML requires a large number of images to train computer models successfully. Although there are medical image datasets available, more large image datasets are needed from a variety of lesions.[6] To address this necessity, we created an image dataset (LC25000) with 25,000 images in 5 classes. Each class contains 5,000 images of the following histologic entities: colon adenocarcinoma, benign colonic tissue, lung adenocarcinoma, lung squamous cell carcinoma and benign lung tissue. All images are de-identified, HIPAA compliant, validated, and freely available for download to AI researchers.[10]

## Acknowledgments

None

## Funding

This material is the result of work supported with resources and the use of facilities at the James A. Haley Veterans' Hospital.